\documentclass[11pt,twocolumn,tighten]{aastex7}

\usepackage{apjfonts}
\usepackage{url}
\usepackage{hyperref}
\usepackage{natbib}
\usepackage{amsmath,amstext,amssymb}
\usepackage[caption=false]{subfig} 
\usepackage{caption}
\DeclareCaptionLabelFormat{moviefmt}{#1~#2\ (Movie)}
\usepackage{xcolor, fontawesome}
\usepackage{color}
\usepackage{enumitem}
\usepackage{rotating} 

\linenumbers 

\newcommand{\kms}{\ensuremath{\rm km\,s^{-1}}}

\renewcommand*{\thefootnote}{\fnsymbol{footnote}}

\newcommand{\carnegie}{Observatories of the Carnegie Institution for Science, Pasadena, CA 91101, USA}
\newcommand{\caltech}{Department of Astronomy, California Institute of Technology, Pasadena, CA 91125, USA}
\newcommand{\standrews}{School of Physics and Astronomy, University of St Andrews, North Haugh, St Andrews, Fife, Scotland KY16 YSS, UK}

\newcommand{\periodhr}{3.930}

\begin{document}

\title{A Plasma Torus Around a Young Low-Mass Star}

\correspondingauthor{Luke G. Bouma}

\received{23 April 2025}
\revised{29 May 2025}
\accepted{6 June 2025}
\shorttitle{CPV Plasma Torus} 

\shortauthors{Bouma et al.}

\author[orcid=0000-0002-0514-5538,sname='Bouma']{Luke~G.~Bouma}
\altaffiliation{Carnegie Fellow; 51 Pegasi b Fellow}
\affiliation{\carnegie}
\affiliation{\caltech}
\email{lbouma@carnegiescience.edu}

\author[orcid=0000-0002-1466-5236,sname='Jardine']{Moira~M.~Jardine}
\affiliation{\standrews}
\email{mmj@st-andrews.ac.uk}


\begin{abstract}
  A small fraction of red dwarfs younger than $\sim$100 million years
  show structured, periodic optical light curves suggestive of
  transiting opaque material that corotates with the star.  However,
  the composition, origin, and even the existence of this material are
  uncertain. The main alternative hypothesis is that these complex
  periodic variables (CPVs) are explained by complex distributions of
  bright or dark regions on the stellar surfaces.  Here, we present
  time-series spectroscopy and photometry of a rapidly-rotating
  ($P$=3.9\,hr) CPV, TIC\,141146667.  The spectra show sinusoidal
  time-varying H$\alpha$ emission at twice to four times the star's
  equatorial velocity, providing direct evidence for cool
  ($\lesssim$10$^4$\,K) plasma clumps trapped in corotation around a
  CPV.  These data support the idea that young, rapidly-rotating M
  dwarfs can sustain warped tori of cool plasma, similar to other
  rapidly-rotating magnetic stars.  Outstanding questions include
  whether dust clumps in these plasma tori explain CPV light curves,
  and whether the tori originate from the star or are fed by external
  sources.  Rough estimates suggest $\gtrsim$10\% of M dwarfs host
  similar structures during their early lives.
\end{abstract}



\section{Introduction}
\label{sec:intro}

Stars with masses below about half that of the Sun, M dwarfs, are the
only type of star to offer near-term prospects for detecting the
atmospheres of rocky exoplanets with surface water.  Community
investment with JWST is proceeding accordingly
\citep[][]{Redfield2024,TRAPPIST1JWSTCommunityInitiative2024}.  The
question of how M dwarfs influence their planets---especially the
retention of their atmospheres---has correspondingly grown in
importance.  Previous work has established that most M dwarfs host
close-in planets \citep{Dressing2015} that on average are subject to
long disk lifetimes \citep{Ribas2015}, intense UV radiation
\citep{France2016}, and frequent stellar flares \citep{Feinstein2020}.
However, despite extensive work in these areas, the plasma and
magnetospheric environments that bathe young, close-in exoplanets
remain challenging to quantify.  Understanding these environments is
crucial because they directly impact atmospheric retention and,
ultimately, habitability.

\begin{figure}[!t]
  \centering
  \includegraphics[width=0.47\textwidth]{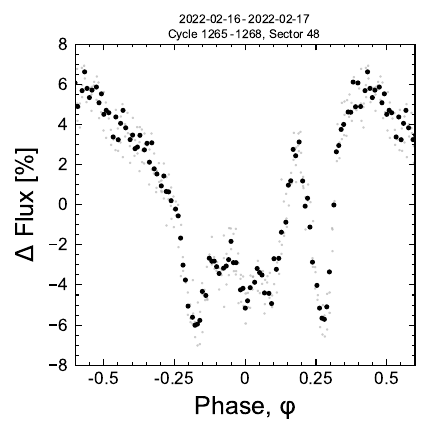}
  \vspace{-0.4cm}
  \captionsetup{labelformat=moviefmt,labelsep=colon}
	\caption{\textbf{TIC\,141146667 is a complex periodic variable (CPV).}  The
online movie,
  \href{https://lgbouma.com/movies/TIC141146667_20250116.mp4}{{\bf
  available here}},
  covers a baseline of 5{,}784 cycles irregularly sampled over three
  years.  The TESS light curve is phased to the \periodhr\ hour period
  in groups of three cycles per frame.  This is the period both of
  stellar rotation, and (we hypothesize) of corotating clumps of
  circumstellar material.  Raw data acquired at two minute cadence are
  in gray; black averages to 100 points per cycle.  The sharp
  photometric features persist for tens to thousands of rotational
  cycles. }
  \label{fig:lc}
\end{figure}

One example of our current ignorance is the complex periodic
variables.  While Figure~\ref{fig:lc} highlights the main object of
interest in this article, over one hundred analogous systems have now
been found by K2 and TESS
\citep{Rebull2016,Stauffer2017,Rebull2018,Zhan2019,Rebull2020,Stauffer2021,Popinchalk2023,Bouma2024}.
These CPVs are phenomenologically identified based on their
structured, periodic optical light curves; most are M dwarfs with
rotation periods shorter than two days.  Within current sensitivity
limits, none host disks \citep{Stauffer2017,Bouma2024}.  However,
$\approx$3\% of stars a few million years old show this complex
behavior, an observed fraction which decreases to $\approx$0.3\% by
$\approx$110\,Myr \citep{Rebull2020}.  CPVs can and have been confused
for transiting exoplanets
\citep{vanEyken2012,Johns-Krull2016,Bouma2020}.

The two leading hypotheses for explaining CPVs are either that
transiting clumps of circumstellar material corotate with the star
\citep{Stauffer2017}, or that these stars represent an extreme in
naturally-occurring distributions of starspots or faculae
\citep{Koen2021}.  The main argument against a starspot-only
explanation invokes the timescales and amplitudes of the sharpest
photometric features.  However, no independent evidence has yet been
acquired for the presence of any circumstellar material.  If such
material exists, then the geometric correction from the transit
probability would imply an intrinsic occurrence rate at least a few
times larger than the observed rate, suggesting that these clumps
could exist around $\gtrsim$10\% of M dwarfs during their early lives.

The dearth of evidence for circumstellar material around CPVs is
surprising given that separate studies of young stars have, for
decades, reported that stellar coronae contain both hot
($\gtrsim$$10^6$ K) and cool ($\lesssim$$10^4$ K) plasma. In
particular, time-series spectroscopy of stars with a wide range of
masses has shown periodic high-velocity absorption and emission in
Balmer lines such as H$\alpha$, interpreted as long-lived, corotating
clumps of cool plasma
\citep{CollierCameron1989,CollierCameron1992,Barnes2000,Donati2000,Dunstone2006,Skelly2008,Leitzinger2016,Cang2021}.
Such clumps are thought to be forced into corotation by the star's
magnetic field, and the exact geometry of where the plasma can
accumulate is dictated by the field's topology \citep{Waugh2022}.  For
instance, a magnetic dipole field tilted with respect to the stellar
spin axis yields accumulations in a warped torus geometry
\citep{Townsend2005}.  To date, none of these spectroscopic variables
have shown any photometric anomalies \citep{Bouma2024}, leaving open
the issue of whether they are related to CPVs.

In this study, we present the first spectroscopic detection of
corotating clumps of cool plasma around a CPV, TIC\,141146667.
Section~\ref{sec:obs} describes our observations;
Section~\ref{sec:results} presents the results; Section~\ref{sec:disc}
discusses their interpretation and highlights future directions.

\section{Observations}
\label{sec:obs}

\begin{table}
\small
\setlength{\tabcolsep}{2pt}
\centering
  \caption{Selected properties of TIC\,141146667.}
\vspace{-0.2cm}
\label{tab:params}
\begin{tabular}{llcc}
\hline \hline
Parameter & Description & Value & Source\\
\hline 
$T_{\rm eff}$\dotfill                   & Effective Temperature (K) \hspace{9pt}\dotfill                 & 2972 $\pm$ 40    & 1 \\
$R_\star$\dotfill                       & Stellar radius ($R_\odot$)\dotfill                             & 0.42$\pm$0.02    & 1 \\
Age                                     & Stellar age range (Myr)\dotfill                                & 35-150           & 2 \\
$M_\star$\dotfill                       & Stellar mass ($M_\odot$)\dotfill                               & 0.22$\pm$0.02    & 3 \\
$\gamma$\dotfill                        & Systemic radial velocity (\kms)\dotfill                        & 0.61 $\pm$ 1.47  & 4 \\
SpT\dotfill                             & Spectral Type\dotfill                                          & M5.5Ve           & 4 \\
$P_{\rm rot}$\dotfill                   & Photometric rotation period (hr)\dotfill                       & $3.930\pm 0.001$ & 5 \\
$v_{\rm eq}$\dotfill		                & Equatorial velocity \dotfill                                   &  130$\pm$4       & 6 \\
                                        & \hspace{3pt} ($2\pi R_\star/P_{\rm rot}$) (\kms)	             &                      \\
$v_{\rm eq}\sin{i_\star}$\dotfill		    & Projected rotational\dotfill                                   &  138$\pm$8       & 4 \\
                                        & \hspace{3pt} velocity (\kms)	                                 &                      \\
$v_{\rm crit}$\dotfill		              & Critical velocity \dotfill                                      &  316$\pm$16      & 6 \\
                                        & \hspace{3pt} ($G M_\star / R_\star$)$^{1/2}$ (\kms)	           &                      \\
$i_\star$\dotfill                       & Stellar inclination\dotfill                                    & 	$>$63           & 4 \\
                                        & \hspace{3pt}  2$\sigma$ lower limit (deg)	                     &                      \\
$d$\dotfill                             & Distance (pc)\dotfill                                          & $57.54 \pm 0.09$ & 7 \\
$R_{\rm c}$\dotfill		                  & Keplerian corotation\dotfill                                   & $1.82 \pm 0.10$  & 6 \\
                                        & \hspace{3pt} radius ($R_\star$)	                               &                      \\
$a_0$\dotfill                           & Mean inner clump (0)\hspace{9pt}\dotfill                       &  2.07$\pm$0.04   & 4 \\
                                        & \hspace{3pt} orbital radius ($R_\star$)	                       &                      \\
$a_1$\dotfill                           & Mean inner clump (1)\hspace{9pt}\dotfill                       &  2.88$\pm$0.10   & 4 \\
                                        & \hspace{3pt} orbital radius ($R_\star$)	                       &                      \\
$a_2$\dotfill                           & Mean outer clump\hspace{9pt}\dotfill                           &  3.88$\pm$0.25   & 4 \\
                                        & \hspace{3pt} orbital radius ($R_\star$)	                       &                      \\
$\langle$EW$_{\rm H\alpha}$$\rangle$\dotfill
                                        & Time-averaged H$\alpha$ line core\dotfill                      &  7.2 $\pm$ 0.2   & 4 \\ 
                                        & \hspace{3pt} equivalent width (\AA)	                           &                      \\
\hline
\end{tabular}
\begin{flushleft}
\footnotesize{ \textsc{NOTE}---
Provenances are:
1: SED fit \citep{Bouma2024}.
2: Gaia DR3 photometry shows the star is on the pre-main sequence,
   while the spectrum lacks lithium (Appendix~\ref{sec:stparams}).
3: PARSEC v1.2S \citep{Chen2014}.
4: Keck/HIRES (Appendix~\ref{subsec:halpha}).
5: TESS light curve.
6: Derived quantity.
7: Gaia DR3 geometric \citep{GaiaDR3}.
}
\end{flushleft}
\vspace{-0.5cm}
\end{table}

\begin{figure*}[!t]
  \centering
  \includegraphics[width=0.925\textwidth]{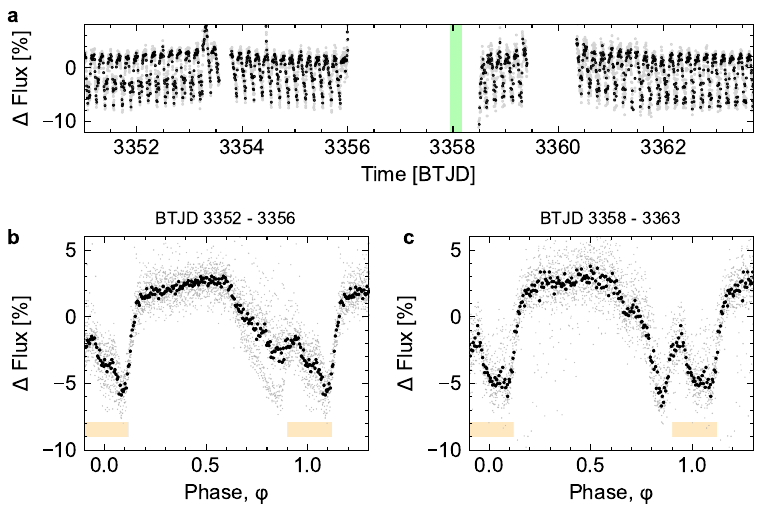}
  \vspace{-0.2cm}
  \caption{{\bf Photometric evolution of TIC\,141146667 around the
  Keck/HIRES observation (green bar).}  {\bf a,} TESS simple aperture
  photometry.  Data gaps were caused by stray light from the Earth
  (BTJD 3356-3358.5) and Moon (BTJD 3359.5-3360.5).  Raw two minute
  data are in gray; black time-averages to ten minute sampling.  {\bf
  b-c,} Folded TESS light curve before and after spectroscopy.  Black
  now phase-averages to 100 points per 3.93\ hour cycle.  During BTJD
  3352-3356, a state switch occurred near BTJD 3353, and the dip at
  $\phi$$\approx$0.8 became less pronounced.  While the dip spanning
  $\phi$=0.6-1.1 was present both before and after the HIRES sequence,
  its photometric shape evolved during the data gap.  The orange bar
  denotes times of spectroscopic transits for the inner two H$\alpha$
  clumps observed with HIRES (see Figure~\ref{fig:spec}).}
  \label{fig:fulllc}
\end{figure*}

\begin{figure*}[!t]
  \centering
  \includegraphics[width=0.99\textwidth]{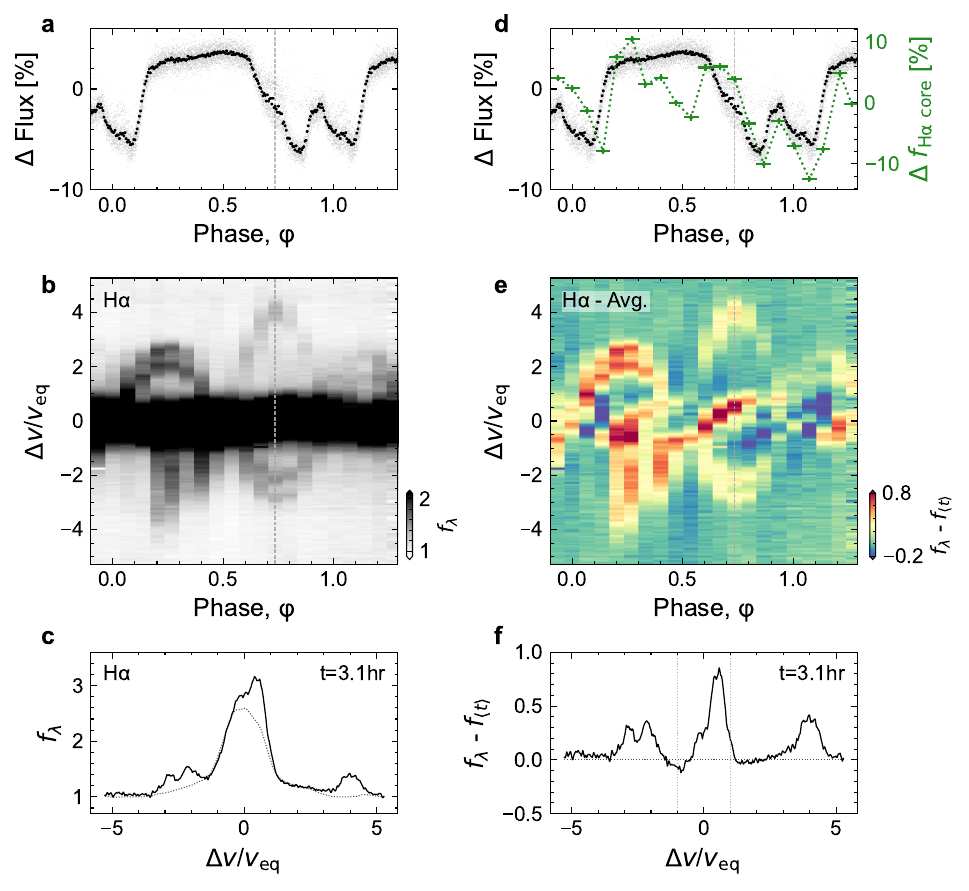}
  \vspace{-0.3cm}
  \captionsetup{labelformat=moviefmt,labelsep=colon}
  \caption{\textbf{Emission from plasma clumps orbiting
  TIC\,141146667.}
  The online movie,
  \href{https://lgbouma.com/movies/TIC141146667_sixpanel.mp4}{{\bf
  available here}}, shows the spectral evolution over five hours.
  {\bf a,} TESS light curve from 5 February 2024 to 26 February 2024
  folded on the 3.93\ hour period.  Black points are phase-averaged;
  gray are the raw data.
  {\bf b,} Keck/HIRES H$\alpha$ spectra from 17 February 2024.  The
  continuum is set to unity, and the darkest color is at twice the
  continuum to accentuate emission outside the line core ($|v/v_{\rm
  eq}|>1$, for the equatorial velocity $v_{\rm eq}$=130\,\kms).  While
  emission within the line core comes from the star's chromosphere,
  the sinusoidal emission features are most readily described by a
  warped plasma torus.
  {\bf c,} Individual epochs of Panel b, visible in the online movie.
  The dotted line shows the time-averaged spectrum, $f_{\langle t
  \rangle}$.
  {\bf d,} As in Panel a, but overplotting the median-normalized
  H$\alpha$ light curve at $|v/v_{\rm eq}|<1$.
  {\bf e,} As in Panel b, after subtracting $f_{\langle t \rangle}$.
  The line core shows H$\alpha$ excesses and decrements advancing from
  the blue to red wings.  The asymmetric color stretch mirrors the
  dynamic range of the data.
  {\bf f,} Individual epochs of Panel e, visible in the online movie.
  A separate version with best-fit sinusoids from
  Appendix~\ref{subsec:halpha} overplotted is also
  \href{https://lgbouma.com/movies/TIC141146667_sixpanel_sinusoids.mp4}{{\bf
  available}}.}
  \label{fig:spec}
\end{figure*}

We identified TIC\,141146667 in previous work \citep{Bouma2024} by
searching TESS two-minute data from 2018-2022 for highly structured,
periodic light curves \citep{Ricker2015}.  We chose the star for
spectroscopy because its brightness and rapid rotation enabled an
efficient search for variability in its line profiles.  As an
apparently single pre-main sequence M dwarf, its properties are
typical for the CPV population (see Table~\ref{tab:params} and
Appendix~\ref{sec:stparams}).

We observed TIC\,141146667 ($V$=16.2) for five hours on
17~February~2024 using the High Resolution Echelle Spectrometer
(HIRES; \citealt{vogt_hires_1994}) on the 10\,m Keck I telescope.  The
observations spanned the second half of the night, from 10:47 to 16:13
(UT).  We opted for a fixed 15 minute cadence, and used the C2 decker
(0$\farcs$86$\times$14$\farcs$0) in the red configuration, yielding a
resolution $R$$\approx$45{,}000 ($\delta v$$\approx$6.7\,\kms).
Strong winds contributed to 1$''$-1\farcs5 seeing, but conditions were
otherwise favorable.  We reduced the echelleogram using the standard
techniques of the California Planet Survey \citep{Howard2010}.  

TESS observed TIC\,141146667 ($T$=13.3) for six non-contiguous months
spanning 2019-2024.  TESS acquired these observations at two-minute
cadence during Sectors 41, 48, and 75.   The 30-minute data during
Sectors 14, 15, and 21 smeared sharp features over the star's
3.93~hour period (see \citealt{Gunther2022}).  The movie in
Figure~\ref{fig:lc} shows the two-minute data: like other CPVs,
TIC\,141146667 maintains a fixed period over timescales of years while
its detailed photometric morphology evolves.  The nearest known star,
TIC\,141146666 ($T$=14.5), is 25$''$ from TIC\,141146667 and is
photometrically quiet; crowding is not a concern.

We observed with Keck during Sector~75 in an attempt to obtain
simultaneous observations.   Figure~\ref{fig:fulllc} shows the result;
Earth passed within 25$^\circ$ of the TESS camera's boresight from
BTJD 3356.0-3358.5, which caused a data gap that ended twelve hours
after our Keck/HIRES observations (green bar).  From BTJD
3359.4-3362.0, the Moon then passed within 25$^\circ$ of the camera's
boresight.  Based on the level of scattered light in the optimal
aperture \citep{Jenkins2016}, we manually masked out data in the TESS
light curve from 3359.4-3360.1; the remainder of the data during the
lunar approach were usable.  Small gaps from BTJD 3353.6-3353.8 and
3360.1-3360.3 were caused by data downlinks at the spacecraft's
perigee and apogee.

While the data gap is unfortunate, Figure~\ref{fig:fulllc} shows that
both before and after the HIRES data were acquired, a large flux
decrement spanned roughly half of each cycle.  From BTJD 3352-3356,
this dip had two sharp local minima;  the minimum at
$\phi$$\approx$0.8 became less pronounced following the flare at BTJD
3353, yielding a dip more closely resembling an asymmetric ``V'' than
a ``W''.  Similar CPV state changes have been previously seen
\citep{Stauffer2017,Bouma2024}.  The photometric shape therefore
evolved during the twelve cycles spanning the 3356-3358 gap, since the
average shape from 3358-3363 more closely resembles the initial ``W''.
Nonetheless, the global photometric morphology---a small brightening
over 45\% of the period, followed by a complex flux dip spanning 55\%
of the period---is similar before and after the data gap.

\section{Results}
\label{sec:results}

Figure~\ref{fig:spec} shows the TESS and HIRES data from February
2024.  The spectra (Panels b,c) show emission in H$\alpha$ beyond the
star's equatorial velocity, $v_{\rm eq}$, of 130\,\kms.  There are at
least two distinct emission components, 180$^\circ$ apart in phase.
The inner component at lower velocities has clearer sinusoidal
behaviour in time and is double-peaked, with peak semi-amplitudes of
2.07\,$v_{\rm eq}$ and 2.88\,$v_{\rm eq}$ (see
Appendix~\ref{subsec:halpha}).  There is significant non-periodic
variability in the emissivity of this double-peaked component: the
flux excess begins with an amplitude 70\% above the continuum, and
diminishes to 10\% by sunrise.  The higher-velocity component
180$^\circ$ opposite in phase is detected only from $\phi$=0.2-1.0.
From $\phi$=0.2-0.5, this outer component appears connected to the
star in velocity space.  While its peak semi-amplitude of
3.88\,$v_{\rm eq}$ is achieved at both $\phi$=0.25 and $\phi$=0.75,
its amplitude similarly decreases from a 60\% excess to a 10\% excess
over the observation sequence.  In Appendix~\ref{subsec:halpha}, we
measured the period for all three emission components to be consistent
with the photometric \periodhr\ hour period, to within two minutes for
the inner component.  

These sinusoidal emission features require circumstellar clumps of
partially-ionized hydrogen to corotate with the star.  Based on the
observed sinusoidal periods and velocities, this material is not
moving on a Keplerian orbit; it is magnetically forced to co-rotate
with the star.  The velocity semi-amplitude of the sinusoids gives the
mean distance of each clump from the stellar rotation axis:
2.07\,$R_\star$ and 2.88\,$R_\star$ for the inner clumps, and
3.88\,$R_\star$ for the outer clump.   These clumps transit in front
of the star when passing from negative to positive velocity.  The
transits of the two inner clumps last $\approx$22\% of each cycle,
from $\phi$=-0.1 to $\phi$=+0.12.  This spectroscopic transit
coincides with the latter half of the complex eclipse feature in the
TESS data (see Figure~\ref{fig:fulllc} and Panels a and b of
Figure~\ref{fig:spec}).

The H$\alpha$ line core, visible in Panels e and f of
Figure~\ref{fig:spec}, is more complex.  At $|\Delta v / v_{\rm
eq}|<1$, most observed H$\alpha$ photons come from the star's
chromosphere; variability in the stellar emission line can be caused
by circumstellar material, bright regions, or dark regions crossing
the star's surface.  Figure~\ref{fig:spec}e suggests that all three
effects occur in TIC\,141146667.  For instance, from $\phi$=0-0.3,
double-peaked emission is visible both on and off-limb; this feature
is circumstellar in origin.  However, the emission feature that
crosses the star from $\phi$=0.4-0.9 is brighter than the
circumstellar clumps, and its slow speed instead suggests it is
associated with a bright region on the star's surface.  Similarly,
from $\phi$=0.6-1.15 a 20\% deep absorption feature slowly crosses the
H$\alpha$ line profile.  This feature could be either a
chromospherically dark region (e.g.~a spot group), or an azimuthally
extended component of the circumstellar material.  The origin of the
other bright and dark streaks passing across the line core in
Figure~\ref{fig:spec}e is similarly ambiguous. 

Figure~\ref{fig:spec}d shows a final exercise to quantify the behavior
of the line core, by summing the H$\alpha$ flux at $|\Delta v / v_{\rm
eq}|<1$.  This panel shows that changes in the line core flux ($f_{\rm
H\alpha\ core}$) correlate with the broadband variability throughout
most of the light curve, except near $\phi$$\approx$0.5, corresponding
to the transit of the 3.9\,$R_\star$ clump and the occultation of the
lower-velocity clump.  Spectroscopic transits may therefore not always
be associated with photometric transits.

\section{Discussion}
\label{sec:disc}

\subsection{Physical Properties of the Emitting Region}


Our Keck/HIRES observations are the first reported time-series spectra
of a CPV, and they demonstrate that corotating circumstellar plasma
clumps exist around at least one such star.  More specifically, the
spectra require plasma with a significant population of hydrogen in
the $n$=3 excited state, with minimal evidence for higher-order
excitations.  Most of this H$\alpha$ emission originates in ``clumps''
with size comparable to the star; radial ``spokes'' or azimuthally
extended ``arcs'' for the emitting material are ruled out by the
$\approx$30\,\kms\ H$\alpha$ velocity dispersion (see
Appendix~\ref{subsec:halpha}).  The H$\alpha$ line luminosity suggests
characteristic number densities and masses for the gaseous component
of these clumps of $n_{\rm H} \sim 10^{11}$\,cm$^{-3}$ and $M_{\rm
gas} \sim 10^{17}$\,g (see Appendix~\ref{subsec:gas}).  Dust is
independently constrained to have a total mass $M_{\rm dust} <
10^{17}\,{\rm g}$ based on the lack of a WISE infrared excess; if one
assumes that the opacity in the TESS flux dips comes from dust, a
lower limit $M_{\rm dust} > 10^{15}\,{\rm g}$ follows (see
Appendix~\ref{subsec:dust}).

\subsection{Clumps Within Warped Plasma Tori}

While the H$\alpha$ emission mostly comes from clumps, the broadband
flux dip in Figure~\ref{fig:spec} spans roughly half of each cycle.
In CPVs more generally, individual flux dips last 5-50\% of each
cycle, and there are often multiple dips per cycle \citep{Bouma2024}.
This implies that azimuthally-distributed material may be the norm for
CPVs; a warped torus may be a more accurate picture than a clump.

\citet{Townsend2005} outlined the physics of how rapidly-rotating
stars with strong magnetic fields can support plasma tori \citep[see
also][]{Nakajima1985,Ferreira2000,Petit2013,Daley-Yates2024}.  When
the magnetospheric radii $R_{\rm m}$ of such stars exceeds their
Keplerian corotation radii $R_{\rm c}$, the effective potential
(gravitational plus centrifugal) experienced by plasma along any given
field line has a local minimum outside $R_{\rm c}$, which enables
charged material to accumulate in a torus.  Warps can occur when there
is misalignment between the spin and magnetic axes.  In general, the
regions in these centrifugally-supported magnetospheres with the
densest plasma accumulations need neither transit nor be opaque in
broadband optical light.  

In a follow-up study, \citet{Townsend2008} made synthetic light
curves, assuming that the optical depth scaled linearly with plasma
density.  They found that W-shaped eclipses, similar to those seen for
TIC\,141146667, can occur when the spin and magnetic axes are
moderately (15-45$^\circ$) misaligned.  The movies associated with
their work are available
online\footnote{\url{http://user.astro.wisc.edu/\~townsend/static.php?ref=rrm-movies\#Download\_Bundles}
last accessed 28 May 2025.}; for a $\omega/\omega_{\rm c}$$\approx$0.5
edge-on system like TIC\,141146667, the photometric eclipses have the
correct shape, and the H$\alpha$ emission similarly exhibits
double-peaked behavior.  The plasma clumps in this model are
180$^\circ$ apart because the deepest local minima in the effective
potential exist along the line of intersection between the rotational
and magnetic planes (see Equation~22 of \citealt{Townsend2005}).  

To summarize, the evidence for warped plasma tori in TIC\,141146667
and CPVs more broadly is that Figure~\ref{fig:spec} shows two plasma
clumps separated by 180$^\circ$ in phase; the warped-torus model
predicts both ``W'' shaped photometric eclipses and double-peaked
H$\alpha$ morphology at quadrature; and CPV dips last 5-50\% of each
cycle, indicating an azimuthally extended distribution of material.
However, two challenges remain. First, the model predicts {\it two}
W-shaped eclipses per cycle.  Second, it predicts that the antipodal
clumps should lie at equal distances from the star.  Different clump
opacities or non-dipolar magnetic fields might resolve these
discrepancies.  Overall, the evidence suggests a warped torus is a
good first approximation, but one that will need refinement in future
work.

\subsection{Origin of CPV Photometric Variability and Astrophysical Analogs}

Microphysically, it is not obvious whether hydrogen alone can produce
the chromatic broadband flux variations seen in CPVs.  The alternative
is that charged dust could provide most of the opacity
\citep{Sanderson2023}.  For TIC\,141146667, Figures~\ref{fig:fulllc}
and~\ref{fig:spec} show that the transits of the inner H$\alpha$ clump
only partially overlap the complex photometric modulation.  This
implies that some of the complex photometric dips must involve
additional opacity sources or spatially distinct structures.

Independent of this opacity question, our observations show that
corotating clumps of cool plasma exist around a CPV.  While starspots
do contribute smooth signals to CPV photometric variability, the
existence of these clumps  would not be predicted by a
``starspot-only'' scenario \citep{Koen2021}.  Scenarios in which the
circumstellar material is made only of dust are similarly ruled out.  

The circumstellar material -- either pure plasma or dusty plasma --
could originate either from the star or from an external source.
Plausible external sources include an undetected old disk, comets, or
a close-in exoplanet.  This latter scenario would make CPVs extrasolar
analogs of the Jupiter-Io torus (e.g.~\citealt{Bagenal1981},
\citealt{Kislyakova2018}), although with a very different composition.

The other CPV analog is the $\sigma$~Ori~E variables, a rare subset of
B stars with radiatively-driven winds that accumulate into warped
plasma tori \citep{Townsend2005}.  These tori tend to have dense
antipodal accumulations of plasma sculpted by tilted-dipole magnetic
fields; these clumps produce broadband optical variability through
bound-free scattering \citep{Townsend2005} and Thomson scattering
\citep{Berry2022}.  For $\sigma$~Ori~E and most of its analogs, the
result is simple light curves that resemble eclipsing binaries, and
time-dynamic H$\alpha$ spectra similar to Figure~\ref{fig:spec}.  The
few known exceptions---HD~37776, HD~64740, and HD~176582---have
complex light curves resembling CPVs \citep{Mikulasek2020,Bouma2024}
and spectropolarimetric measurements that may suggest non-dipolar
field contributions \citep{Kochukhov2011,Shultz2018}.  The photometric
complexity of CPVs may therefore be related to magnetic fields with
highly multipolar contributions.  While this suggestion is consistent
with the wide range of axisymmetric and non-axisymmetric topologies
seen in mid- and late-M dwarfs
(e.g.~\citealt{Donati2006,Kochukhov2017,Shulyak2019,Bellotti2024}),
future work is needed to verify whether CPVs indeed select for
complex, non-axisymmetric fields.

\subsection{Connection to Previous Work}

Spectra of magnetically-active, rapidly rotating stars with a wide
range of masses have been previously observed to exhibit sinusoidal
time-varying Balmer emission
\citep{Donati2000,Townsend2005b,Dunstone2006,Skelly2008}, similar to
Figure~\ref{fig:spec}.  No such stars were previously known to show
complex light curves \citep{Bouma2024}.   One interpretation for the
spectroscopic variability of these stars, and that of the analogous
transient H$\alpha$ absorption features
\citep{CollierCameron1989,CollierCameron1992,Cang2020}, comes from an
analogy to quiescent solar prominences, cool condensations of plasma
in the solar corona that can last days to weeks
\citep{VialEngvold2015}.  In the case of the Sun, these condensations
fall back to the photosphere because gravity is stronger than any
magnetic tension or centrifugal force capable of sustaining them.
However, it has been understood at least since work by
\citet{Donati2000} that these ``prominence systems'' can be
longitudinally extended, forming a trapped ring of plasma.  While the
issue of why these previously known spectroscopic variables do not
show complex photometric variability remains open, plausible
explanations include that they are not in the required transiting
geometry, or that they lack the necessary source of opacity.  What we
have in the case of TIC\,141146667 is a system that finally exhibits
both sets of phenomena.

\subsection{Future Work: Composition, Origin, Modeling}

Pressing issues include determining the composition and origin of the
circumstellar material, understanding the exact role of the stellar
magnetic field, and exploring the implied space weather experienced by
the close-in rocky exoplanets that, statistically
\citep{Dressing2015}, are likely to be present in many CPV systems.

The material's composition -- either pure plasma or dusty plasma --
can be clarified by time-series infrared spectrophotometry.  While
observations of CPVs in the optical suggest chromaticity consistent
with dust \citep{Tanimoto2020,Gunther2022,Koen2023}, electron
scattering in a plasma transiting over a spotted star might also
produce chromatic features \citep{Rackham2018}.  This degeneracy is
alleviated in the infrared, where the composition and size
distribution of any dust that is present could be determined by
measuring the extinction curve for a sample of CPVs from 1-20\,$\mu$m.
Composition and size distributions similar to debris from rocky bodies
seen around white dwarfs \citep{Reach2009} would suggest an extrinsic
origin channel.  Compositions and sizes similar to the interstellar
medium would suggest that dust can condense out from M dwarf winds,
similar to processes that occur around evolved stars
\citep{Marigo2008}.

Future models could take a range of steps to clarify the origin of
CPVs.  A first step for magnetohydrodynamic models that can currently
explain the H$\alpha$ features \citep[e.g.][]{Waugh2022} would be to
incorporate an opacity source that can reproduce the photometric dips.
This effort will also require exploring what stellar magnetic field
topologies reproduce the observed light curve shapes.  While previous
models in the rigid-field approximation have explored both dipolar
\citep{Townsend2008} and multipolar fields \citep{Krticka2022}, dynamo
simulations of fully-convective M dwarfs have suggested that
global-scale mean fields might be confined to a single hemisphere
\citep{Brown2020}; such fields would yield accumulation surfaces
different from those that have been explored.  One related challenge
is that CPVs currently suffer a dearth of stellar surface field
measurements, unlike benchmark systems such as AB Dor and V374~Peg
\citep{Jardine2002,Vidotto2011}. This might be alleviated by future
spectropolarimetry, or by radio observations sensitive to the magnetic
field structure around the star \citep[][]{Callingham2021,Kaur2024}.
Finally, the question of what microphysical process yields the opaque
material also merits future theoretical attention. While recent
advances can produce the catastrophic cooling that leads to dense,
cool clumps embedded in a hot corona
\citep{Daley-Yates2023,Daley-Yates2024}, a self-consistent treatment
of radiation propagating through such clumps has yet to be performed.
Whether dust grains can form in these plasma clumps similarly deserves
attention.

It is currently unclear what, if any, relationship CPVs have to the
close-in rocky exoplanets that exist around most M dwarfs
\citep{Dressing2015}.  However, a few percent of young M dwarfs show
the CPV phenomenon \citep{Rebull2020}, and our data show that some of
the flux dips occur when clumps of circumstellar material transit the
star.  The geometric correction implies that an appreciable minority
($\gtrsim$10\%) of young M dwarfs---the rapidly rotating ones with
centrifugal magnetospheres---host circumstellar environments similar
to the CPVs.  Future studies that combine spectroscopic, polarimetric,
and multi-wavelength observations, along with magnetohydrodynamic
modeling, will be key to understanding the complex environments of
these young stars.

\begin{acknowledgements}
  We thank B.~Tofflemire, A.~Weinberger, and L.~Hillenbrand for
  conversations that significantly informed this work; J.~Spake,
  J.~Winn, J.-F~Donati, R.~Townsend, and O.~Kochukhov for feedback on
  the manuscript; and A.~Howard and H.~Isaacson for their assistance
  reducing the HIRES spectra.

  LGB acknowledges support from the Carnegie Fellowship and the
  Heising-Simons 51~Pegasi~b Fellowship.
  MMJ acknowledges support from STFC consolidated grant number
  ST/R000824/1.

  The analyses in this article used data from TESS \citep{TESS2min}.
  The 2-minute cadence observations from Sectors 41 \& 48 were from
  TESS DDT039 (PI: M.~Kunimoto) and the observations from Sector 75
  were from TESS G06030 (PI: L.~Bouma).
  Funding for the TESS mission is provided by NASA's Science Mission
  directorate.
  TESS is a product of millions of hours of work by thousands of
  people, and we thank the TESS team for their efforts to make the
  mission a continued success.  The HIRES data were obtained at the
  Keck Observatory, which exists through a similar scale of community
  effort.
  We recognize the importance that the summit of Maunakea has always
  had within the indigenous Hawaiian community, and we are grateful
  for the opportunity to conduct observations from this mountain.
\end{acknowledgements}

{\it \large Contributions}: 
LGB led the inception, design, execution, and interpretation of the
project.  MMJ contributed to the interpretation.

\facilities{
  Gaia \citep{GaiaDR3},
  TESS \citep{Ricker2015},
  Keck:I (HIRES) \citep{vogt_hires_1994},
	2MASS \citep{Skrutskie2006},
	SDSS \citep{2000AJ....120.1579Y}.
}

\software{
  astroariadne \citep{Vines2022},
  astropy \citep{astropy:2013,astropy:2018,astropy:2022},
  chatgpt \citep{chatgpt2025openai},
	dustmaps \citep{2018JOSS....3..695M},
  matplotlib \citep{matplotlib},
  numpy \citep{numpy},
  numpyro \citep{Phan2019},
  scipy \citep{scipy}.
}


\appendix

\section{Stellar Parameters}
\label{sec:stparams}

{\it Radial Velocity}---We measured the radial velocities of
TIC\,141146667 (Gaia DR3 860453786736413568) from our HIRES spectra
using a custom pipeline.  Our method was based on template-matching
against synthetic PHOENIX spectra \citep{Husser2013}, and was
calibrated against the standard stars described by \citet{Chubak2012}.
We used velocity standards spanning G2-M4, irrespective of rotation
rate.  We included velocity corrections due to Earth's motion around
the solar system barycenter and due to Earth's daily rotation about
its axis, calculated using \texttt{barycorrpy} \citep{Kanodia2018}.
Our analysis code reproduced the systemic velocities of known velocity
standards \citep{Chubak2012} with an RMS of 0.66\,\kms.

For TIC\,141146667, we measured the radial velocities using regions
near the K~I (7700\,\AA) resonance line and three TiO bandheads
(5160\,\AA, 5450\,\AA, and 5600\,\AA).  We selected these regions
because they provided the best matches between the synthetic and
observed spectra.  We averaged the redshift measurements over each
order, and used the scatter of resulting velocity measurements between
orders to assign the RV uncertainty at each epoch.  The
uncertainty-weighted mean systemic velocity over all epochs on
17~February~2024 was $\gamma$=0.6$\pm$1.5\,\kms.  The relative radial
velocities about this mean are listed in Table~\ref{tab:rv}.

{\it Viewing Orientation}---We fitted the rotational broadening of the
K~I (7700\,\AA) resonance line using the kernel suggested by
\citet{Gray2008}; taking the mean and standard deviation of the
resulting value over all epochs yielded $v_{\rm eq} \sin
i$=138$\pm$8\,\kms, consistent with the visual line broadening $\Delta
\lambda$$\approx$3\,\AA.  The star's equatorial velocity $v_{\rm eq}$
based on its apparent size and rotation period is 130$\pm$4\,\kms.
While these measurements are consistent with an edge-on viewing
geometry, the formal constraint is rather weak, with $i$$>$63$^\circ$
at 2$\sigma$ (2.5$^{\rm th}$ percentile of the inclination posterior).


{\it No Evidence For Binarity}---Any periodicity in the radial
velocity time-series is ruled out at the rotation period for
semi-amplitudes above 2.85\,\kms\ (at 3$\sigma$ confidence).  This
sets an upper limit on the mass of any putative companion on a four
hour orbit of $m \sin i $$<$2.4\,$M_{\rm Jup}$.  Regarding possible
companions at wider separations, the Gaia DR3 renormalized unit weight
error (RUWE) is 1.23, within the usual range for apparently single
sources.  There are no resolved companions in the Gaia DR3 point
source catalog.  Finally, we checked the TESS light curve for evidence
of secondary photometric periods by subtracting the mean CPV signal
over each sector and performing a phase-dispersion minimization
analysis \citep{Stellingwerf1978,2021zndo...1011188B}.  There were no
secondary periods in the TESS data.  Previous work \citep{Bouma2024}
has shown that $\approx$30\% of CPVs show evidence for binarity from
RUWE, and $\approx$40\% of CPVs show evidence for binarity based on
the presence of secondary photometric periods.  This agrees with
analyses showing that multi-periodic low-mass stars are usually
unresolved binaries \citep{Tokovinin2018}.  Overall, the CPV binary
fraction seems consistent with that of field M dwarfs
\citep{Winters2019}, pointing to a weak or non-existent connection
between the CPV phenomenon and (wide) binarity.  For TIC\,141146667
specifically, we find no evidence for stellar multiplicity.

\begin{figure}[!t]
  \centering
  \includegraphics[width=0.48\textwidth]{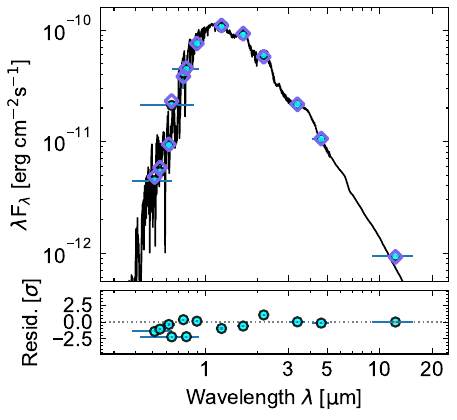}
  \caption{
    {\bf Spectral energy distribution} of broadband photometric
    magnitudes (filled cyan circles) plotted over the best-fit
    BT-Settl stellar atmosphere model \citep{Allard2012} and the
    associated photometric predictions (empty purple diamonds).  This
    plot was made from an adaptation of \texttt{astroARIADNE}
    \citep{Vines2022}.  The photometry extends from the Gaia DR3 blue
    passband to WISE W3; the W4 passband (22 $\mu$m) did not yield a
    confident detection.  This fit yields the star's temperature and
    size.  The lack of excess infrared flux relative to the
    photospheric model sets an upper limit on emission from
    circumstellar dust.
    }
  \label{fig:sed}
\end{figure}

{\it Effective temperature, radius, mass, and spectral
classification}---We adopted the star's effective temperature and
radius measured using the spectral energy distribution (SED) fitting
procedure described by \citet{Bouma2024}.  To summarize, this approach
used \texttt{astroARIADNE} \citep{Vines2022} to fit broadband
magnitudes from Gaia DR2, APASS, 2MASS, SDSS, and WISE $W1$ and $W2$
with the BT-Settl stellar atmosphere models \citep{Allard2012}.  Free
parameters included  the stellar effective temperature, radius,
reddening, surface gravity, and metallicity.  The resulting best-fit
SED is shown in Figure~\ref{fig:sed}.  This method has the most
constraining power for the star's effective temperature (2972 $\pm$
40\,K) and radius (0.42 $\pm$ 0.02\,$R_\odot$).  We measured the
star's spectral type to be M5.5Ve by visually comparing the HIRES
spectra against the photometric standards tabulated by
\citet{Bochanski2007}.   We measured the equivalent width of the
H$\alpha$ line by fitting a range of models to the time-averaged line
profile shown in Figure~\ref{fig:spec}, selecting the model that
minimized the Bayesian information criterion, and numerically
integrating this best fit model.  We found a sum of two Gaussians to
be preferred; our quoted result, EW$_{{\rm
H}\alpha}$=7.2$\pm$0.2\,\AA, comes from numerically integrating within
$|\Delta v/v_{\rm eq}|<1$.  Integrating over the entire line profile,
including the broad wings, would yield EW$_{{\rm
H}\alpha}$=10.2$\pm$0.3\,\AA. Either value would classify the star as
a weak-lined T Tauri \citep{Briceno2019}.

Given the effective temperature, stellar radius, and age range
(35-150\,Myr) derived below, we then estimated the stellar mass by
interpolating against the PARSEC v1.2S isochrones \citep{Chen2014}, as
in \citet{Bouma2024}.  This exercise yielded a mass of
$M_\star$=$0.20\pm0.01$\,$M_\odot$ assuming an age of 35\,Myr, or a
mass of $0.25\pm0.01$\,$M_\odot$ assuming an age of 150\,Myr.  These
masses imply Keplerian corotation radii $R_{\rm
cr}/R_\star$=1.75$\pm$0.07 and $R_{\rm cr}/R_\star$=1.89$\pm$0.07,
respectively; this size scale is expected to set the inner boundary at
which corotating material might accumulate
(e.g.~\citealt{Townsend2005,Daley-Yates2024}).  Our final quoted
$M_\star$ and $R_{\rm cr}$ values adopt the average of these extremes
and a quadrature sum of their statistical uncertainties; a more
precise age would be needed to resolve the systematic uncertainties in
these parameters.

{\it Age: No Obvious Association Membership}---In \citet{Bouma2024} we
previously found that over 90\% of CPVs within 100\,pc are associated
with known young moving groups based on their positions and
kinematics.  TIC\,141146667 is in the minority.  We calculated the
probability of TIC\,141146667 being part of any nearby known group
using BANYAN\,$\Sigma$ v1.2 \citep{Gagne2018}.  That particular model
classifies it as a field star at $>$99.9\% confidence.  We also
searched the local vicinity of TIC\,141146667 for neighbors with
similar projected on-sky velocities using \texttt{comove}
\citep{Tofflemire2021}.  This yielded no strong candidates for
co-moving stars with projected tangential velocities $\Delta v_{\rm
T}$$<$5\,\kms\ that share its isochronal youth.

\begin{figure}[!t]
	\begin{center}
		\subfloat{
			\includegraphics[width=0.46\textwidth]{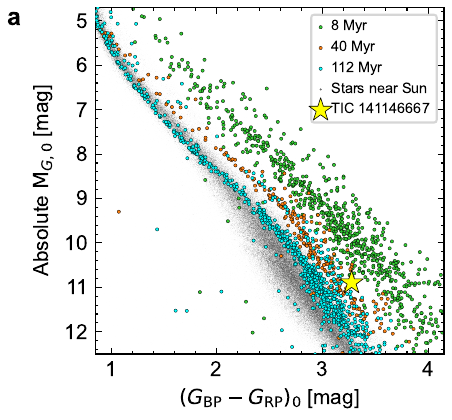}
		}
		\subfloat{
			\includegraphics[width=0.47\textwidth]{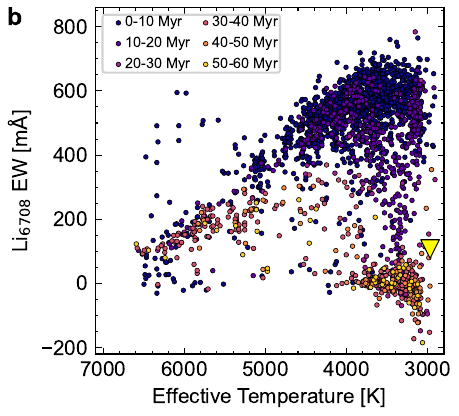}
		}
	\end{center}
	\vspace{-0.3cm}
  \caption{{\bf TIC\,141146667 age diagnostics.}
  {\bf a,} Dereddened Gaia DR3 color vs.~absolute magnitude for
  TIC\,141146667, USco (8\,Myr), IC\,2602 (40\,Myr), the Pleiades
  (112\,Myr) and stars within 100\,pc.  The location of TIC\,141146667
  in this diagram suggests an age of 30-150\,Myr. 
  {\bf b,} A dearth of photospheric lithium for TIC\,141146667 (yellow
  triangle denotes 2$\sigma$ upper limit) yields a lower bound on the
  star's age of $\gtrsim$20\,Myr.  Comparison stars are from the
  Gaia-ESO survey \citep{Jeffries2023}; rich clusters in these data
  include NGC\,2264 (4.5\,Myr), $\lambda$~Ori (8.7\,Myr), ${\rm
  \gamma}^2$\,Vel (16.4\,Myr), NGC\,2547 (35.0\,Myr), IC\,2391
  (42.0\,Myr), and NGC\,2451A (50.0\,Myr). Measurement
  uncertainties are large below $+$50\,m\AA; negative EW does not
  correspond to significant emission.}
  \label{fig:age_diagnostics}
\end{figure}

{\it Age: Isochrones}---The color and absolute magnitude of
TIC\,141146667 suggest that it is a pre-main sequence M dwarf, similar
to all other known CPVs \citep{Stauffer2017,Stauffer2021,Bouma2024}.
The star's proximity ($d$=58\,pc) and high galactic latitude
($b$=$+$53$^\circ$) yield negligible interstellar reddening along the
line of sight \citep{Green2019}.  Figure~\ref{fig:age_diagnostics}
shows the location of TIC\,141146667 in the color--absolute magnitude
diagram (CAMD) relative to young stellar populations including Upper
Scorpius (USco), IC~2602, and the Pleiades.  To make this diagram, we
adopted the USco members in the $\delta$~Sco and $\sigma$~Sco
sub-associations from \citet{Ratzenbock2023}, and the IC~2602 and
Pleiades members from \citet{Hunt2024}.  We assumed an average
$V$-band extinction $A_{\rm V}$=$\{0.12, 0.11, 0.10\}\,{\rm mag}$ for
USco \citep{Pecaut2016}, IC~2602 \citep{Hunt2024}, and the Pleiades
\citep{Hunt2024} respectively, and ages of 8\,Myr
\citep{Ratzenbock2023}, 40\,Myr \citep{Randich2018}, and 112\,Myr
\citep{Dahm2015} for each respective cluster.  We dereddened the
photometry using the extinction coefficients $k_X\equiv A_X/A_0$
tabulated in \citep{GaiaCollaboration2018}, assuming that $A_0 = 3.1
E(B-V)$.

Figure~\ref{fig:age_diagnostics} shows that TIC\,141146667 falls
between the USco and Pleiades sequences, and approximately overlaps
IC~2602.  However, the precision of the implied age is set by the
intrinsic scatter of these calibration sequences; the most luminous
stars in the Pleiades of the same color have a similar absolute
magnitude as TIC\,141146667.  Previous work by \citet{Stauffer2021}
has noted that in the Gaia passbands, CPVs tend to be photometrically
redder and more luminous than single stars in any given cluster; they
also found suggestive evidence for either hotspots or emission from
dust based on analysis of Gaia and WISE photometry.  The implication
is that isochronal age dating of individual field CPVs suffers
significant systematic uncertainty due to factors including dark and
bright spots, magnetic inflation, dust emission, and photometric
binarity. Rather than adopting a particular model isochrone framework,
we instead visually take the star's location in the color--absolute
magnitude diagram to suggest age bounds $t_{\rm
CAMD}$$\sim$30-150\,Myr.

{\it Age: Lithium}---Lithium burning in the cores of low-mass stars
has been studied for over sixty years
\citep[e.g.][]{Hayashi1963,Bildsten1997}.  \citet{Wood2023} provided a
recent overview; the brief summary is that sufficiently cool and young
M dwarfs show the 6707.8\,\AA\ doublet in absorption, $\gtrsim$10\%
below their continua.  However, unlike for Sun-like stars, the
continuum for M dwarfs is challenging to define due to their molecular
absorption.  We therefore attempted a lithium measurement by
constructing a wavelength-binned and Doppler-corrected TIC\,141146667
spectrum, and assigned its uncertainties based on the measured scatter
across the five hour dataset.  We then compared this average spectrum
against the nearest matching M6 template from \citet{Bochanski2007}.
The data show a small depression near the expected lithium wavelength,
potentially consistent with the $\Delta \lambda$$\approx$3\,\AA\ line
broadening.  This feature yields ${\rm EW}_{\rm
Li}$=71$^{+18}_{-13}$\,m\AA, where the statistical uncertainties are
evaluated using a bootstrap resampling technique from the statistical
uncertainties in the HIRES spectrum.  However, the systematic
uncertainty associated with the continuum normalization is likely
comparable to the amplitude of the feature; we therefore treat this
measurement as a $2\sigma$ upper limit: ${\rm EW}_{\rm
Li}$$<$107\,m\AA.

What can be stated with confidence is that lithium is not abundant in
the spectrum of TIC\,141146667.  Figure~\ref{fig:age_diagnostics}
compares our upper limit against the equivalent width measurements
reported by \citet{Jeffries2023} based on the Gaia-ESO spectroscopic
survey.  If the star were $\lesssim$20\,Myr old, at its temperature we
would expect to see lithium in abundance ($>$400\,m\AA).  Since we do
not, we can set an empirical bound on the lithium-derived age of
$t_{\rm Li,emp}$$\gtrsim$20\,Myr.  The \citet{Feiden2016} lithium
isochrones provide a point for theoretical comparison, and suggest
that since $M_{\rm K}$$\approx$6.67\,mag, $t_{\rm
Li,th}$$\gtrsim$35\,Myr is the theoretical age at which complete
depletion occurs in a star with this luminosity (see e.g.~Figure 7
from \citealt{Wood2023}).

{\it Age: Summary}---The main indicators for the youth of
TIC\,141146667 are {\it i)} that it is a complex periodic variable,
and {\it ii)} that it is 1.5 magnitudes brighter (four times more
luminous) than main sequence stars of the same color, while showing no
indicators for binarity.  Being a CPV suggests that the star is young
because a previous CPV search unbiased in age found 90\% of its
detections to be in $\lesssim$200\,Myr old clusters \citep{Bouma2024};
the remaining 10\% were not associated with any coeval population.
Similarly, studies of rotation in $\lesssim$100\,Myr clusters
serendipitously found $\approx$50-100 examples of the class
\citep{Rebull2016,Stauffer2017,Stauffer2018,Rebull2018,Zhan2019,Rebull2020,Stauffer2021,Rebull2022,Popinchalk2023},
whereas analogous studies of Praesepe and the Hyades did not report
any evidence for CPVs in a set of approximately one thousand
$\approx$700\,Myr stars \citep{Rebull2017,Douglas2019,Rampalli2021}.
Regarding the isochronal age constraint, the Pleiades (112\,Myr,
\citealt{Dahm2015}) shows a few stars of equal luminosity and the same
temperature, suggesting a photometric isochronal age upper limit
$\lesssim$150\,Myr.  The weak lithium absorption suggests an age of at
least 20\,Myr based on an empirical comparison using Gaia-ESO spectra,
or at least 35\,Myr based on the \citet{Feiden2016} isochrones.  These
considerations yield our adopted age range of 35-150\,Myr.

\section{Detailed Behavior of H$\alpha$: Model and Implications}
\label{subsec:halpha}

\begin{figure*}[!tp]
  \centering
  \includegraphics[width=0.5\textwidth]{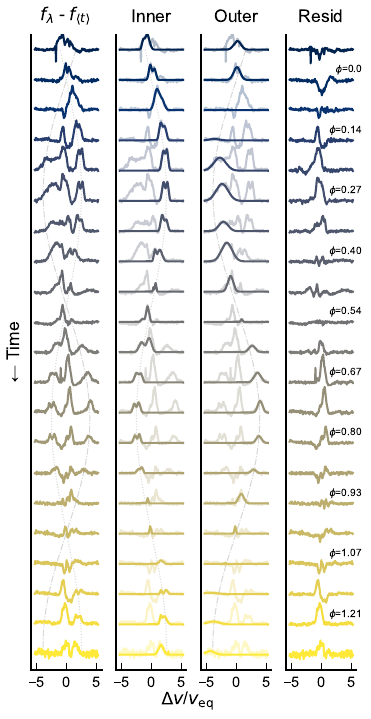}
  \caption{{\bf Time-variable fit to H$\alpha$ line profiles.}  {\it
  Left column:} Raw spectrum at each epoch $f_{\lambda}$ minus the
  time-averaged spectrum $f_{\langle t \rangle}$ (as in
  Figure~\ref{fig:spec}e).  Underplotted sinusoids are not fits; they
  are meant to guide the eye.  {\it Middle columns:} Model of emission
  from the inner clump (sum of two gaussians) and the outer clump
  (single gaussian), plotted over the data.  {\it Right column:}
  Residual of the left column after subtracting the sum of the two
  middle columns, leaving variability in the line core.
  Appendix~\ref{subsec:halpha} discusses the use of this model.  }
  \label{fig:halphamodel}
\end{figure*}

{\it A model for the time-dynamic H$\alpha$ spectrum}---While
Figure~\ref{fig:spec} shows that clumps of circumstellar material
exist around TIC\,141146667, there is value in quantifying the exact
orbital periods, velocities, and velocity dispersions of these clumps.
These quantities can constrain the physical dimensions of the emitting
region, and can also clarify whether the spectroscopic period agrees
with the photometric period.

Given that a full radiative transfer simulation was outside our scope,
we opted to construct a phenomenological model aimed at capturing the
emission from the circumstellar material.  We did this by fitting each
spectral epoch with a multi-component gaussian, after having
subtracted the time-average line profile as in Figure~\ref{fig:spec}e.
The results of this exercise are shown in
Figure~\ref{fig:halphamodel}; details in the implementation and
interpretation follow.

To implement this model, we assumed that the ``inner'' ($K_{\rm
inner}$$\approx$2.5\,$v_{\rm eq}$) clump would be well-fit by a sum of
two gaussians because it is visually double-peaked in the raw data
from $\phi$=0.15-0.35 and $\phi$=0.65-0.85 (Figure~\ref{fig:spec}b).
We assumed that the ``outer'' ($K_{\rm outer}$$\approx$3.9\,$v_{\rm
eq}$) clump would be better fit by a single gaussian, based on its
behavior from $\phi$=0.6-0.9.  Each gaussian component has three free
parameters at each spectral epoch: the mean $\mu$, standard deviation
$\sigma$, and amplitude $A$.  We labeled the inner component's two
gaussians $i=\{ 0, 1 \}$, and the single outer component as $i=2$.
Given the complexity of the line profile data
(Figure~\ref{fig:halphamodel} left column), the likelihood function
for this model is multimodal.  We therefore imposed the prior
constraints that $A_i \sim \mathcal{U}[0,1]$,
$\sigma_i/v_{\rm eq} \sim \mathcal{U}[0,1]$, and further assumed
$\mu_i(t) \sim
\mathcal{U}[
  K_{\rm inner}\sin(\phi(t)) - v_{\rm eq},
  K_{\rm inner}\sin(\phi(t)) + v_{\rm eq}
]$
for the inner two components, and
$\mu_2(t) \sim
\mathcal{U}[
  K_{\rm outer}\sin(\phi(t) + \pi) - 2v_{\rm eq},
  K_{\rm outer}\sin(\phi(t) + \pi) + 2v_{\rm eq}
]$
for the outer component.  This prior on the means mitigates
multimodality in the likelihood by requiring the mean velocity of each
component to be within a one or two $v_{\rm eq}$ of the time-variable
sinusoid suggested by visual inspection.  We fitted each component to
the data {\it independently} using scipy's non-linear least squares
\texttt{curve\_fit} implementation \citep{Virtanen2020}, and scaled
the resulting parameter covariance matrix by a constant factor to
match the sample variance of the residuals.  The middle columns of
Figure~\ref{fig:halphamodel} show the results; a table of the fitted
means, amplitudes, and standard deviations for each component is
available upon request.

Caution is required in interpreting this model's results.  At some
epochs there are no significant spectral features around any
component's prior.  During such epochs, e.g.~the ``outer'' clump at
$\phi$=1.07, the model fits noise, not signal.  At other times, the
model underfits.  For instance, the sudden blue rise near $\phi$=0.2
is poorly described by a gaussian; the assumed functional form is one
of convenience.  Finally, the model fits the $i=\{ 0, 1 \}$, and $i=2$
components independently.  At $\phi$=0.0 and 1.0,
Figure~\ref{fig:halphamodel} suggests that the emission might come
from either the inner or outer components.  Physically however, at
this epoch the inner clump is in transit, and the outer clump is
passing behind the star.  At $\phi$=0.0, the double-peaked emission
profile also matches that seen shortly afterward (at $\phi$=0.14) when
the inner clump is viewed off-disk.  A physical interpretation of the
model would therefore discard the outer clump results at this
particular epoch, because its emission would be blocked by the star.

\begin{figure*}[!t]
  \centering
  \includegraphics[width=0.99\textwidth]{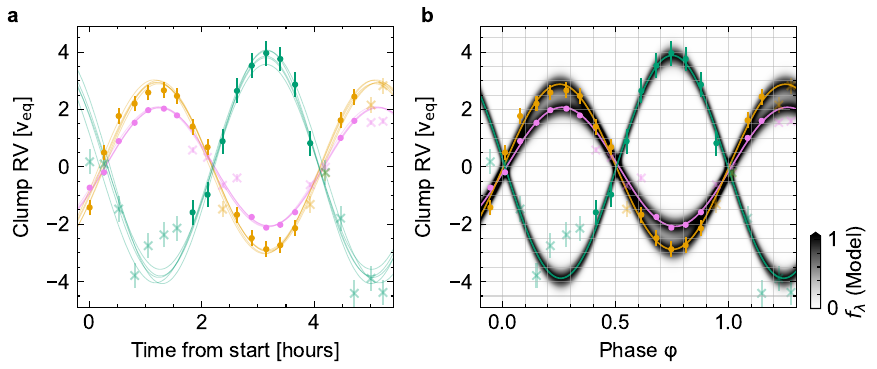}
  \caption{{\bf a,} Orbits fit to mean radial velocities (RVs)
  extracted from H$\alpha$ profile fits in
  Figure~\ref{fig:halphamodel}.  Radial velocities on the vertical
  axis are in units of the equatorial velocity, $v_{\rm
  eq}$=130\,\kms.  Each marker denotes the best-fit gaussian mean at a
  given epoch.  Solid circles were adopted in the fits; transparent X
  markers were excluded due to reliability concerns (see text).  The
  inner double-peaked clump is shown in violet and orange; the outer
  clump is shown in green.  Five model draws from each posterior
  probability distribution are plotted.  {\bf b,} Idealized emissivity
  from mean model fits in Panel a, assuming a constant emission
  amplitude and velocity width $\sigma$=0.25\,$v_{\rm eq}$ and
  neglecting eclipses.}
  \label{fig:orbits}
\end{figure*}

Accounting for these caveats, we used the best-fit parameters from the
multi-gaussian model to quantify the orbital periods and velocity
semi-amplitudes of the clumps.  Figure~\ref{fig:orbits}a shows the
results assuming circular orbits; considering the Bayesian information
criterion, we found no reason to prefer eccentric orbits.  By visually
inspecting Figure~\ref{fig:halphamodel}, we excluded epochs where our
gaussian profile fitting failed to either detect or else adequately
represent the circumstellar emission.  We then used NumPyro to sample
the Gaussian likelihood for a circular orbit with the NUTS algorithm
\citep{Phan2019}.  We used the measurement uncertainties from each
estimated mean radial velocity value and included an additional jitter
term in quadrature.  This procedure yielded orbital periods and
semi-amplitudes of
$P_0$=$3.92\pm0.03$\,hr, $K_0/v_{\rm eq}$=2.07$\pm$0.04;
$P_1$=$3.92\pm0.06$\,hr, $K_1/v_{\rm eq}$=2.88$\pm$0.10;
and $P_2$=$3.88\pm0.20$\,hr, $K_2/v_{\rm eq}$=3.88$\pm$0.25.
The periods for the inner double-peaked clump are therefore consistent
with the photometric $3.930\pm0.001$\,hr period within a precision of
two minutes.  The ``period'' for the outer clump is ambiguous because
the H$\alpha$ data only support the idea of a periodic orbit of
material well-fit by gaussian emission from $\phi$$\approx$0.5-1.0.
From $\phi$=0-0.2, there is no detectable emission, and from
$\phi$=0.2-0.4, the emission spans 1-4\,$v_{\rm eq}$ without taking a
clear gaussian shape.  While the outer-most edge of this
$\phi$=0.2-0.4 emission provides a plausible match to the expectation
of a circular orbit, the idea of invoking a particular functional form
for this component seems fine-tuned.  We instead emphasize that
although this emission is present, its variability in time is
inconsistent with the idea of a stable clump of material.  Additional
observations would be needed to conclusively determine whether or not
this component of the system is long-lived.

For Figure~\ref{fig:orbits}b, we then used the mean orbits from
Figure~\ref{fig:orbits}a to generate sinusoidal-in-time gaussians,
similar to the observations.  We assumed a constant emission amplitude
and velocity width $\sigma$=0.25\,$v_{\rm eq}$ for this exercise, and
normalized each gaussian to unit amplitude;  the colorbar in
Figure~\ref{fig:orbits}b thus masks the non-physical additive
contribution near the zero-crossing of velocity.  Compared to the
behavior of the data at $\phi$=0.2-0.5 (Figure~\ref{fig:spec}), this
is highly idealized.  Nonetheless, this exercise indicates that the
transit of the inner clump lasts $\approx$22\% of each cycle, with a
slight asymmetry around $\phi$=0.

{\it Physical dimensions of emitting region}---The H$\alpha$ velocity
widths in Figure~\ref{fig:halphamodel} constrain the size of the
emitting region via the condition for rigid corotation.  Consider a
clump in cylindrical coordinates with arbitrary radial extent $r$,
azimuthal extent $\ell$, and height $z$.  A range of shapes, including
an ``arc'' with $\ell \gg r$, a ``spoke'' with $r \gg \ell$, and a
``blob'' with $r \approx \ell \approx z$ are all a priori possible.
However, at quadrature, the observed velocity width of emission is
sensitive to the radial extent of the circumstellar material.  At
mid-transit, the observed velocity width is sensitive to the azimuthal
extent.  An arc configuration would minimize the observed velocity
width $\sigma$ at quadrature, and maximize it during transit, with
$\gtrsim$100\,\kms\ variations in between.  Since this is not
observed, the arc geometry, and by a similar argument the spoke
geometry, can be discarded.  

At quadrature, the inner clumps show $\sigma_i$$\approx$0.24\,$v_{\rm
eq}$, implying that 68\% of the emission comes from a volume with
length in the radial dimension $r_i$=$2\sigma_i /
\Omega$$=$0.48\,$R_\star$, and that 95\% of the emission comes from
within 0.96\,$R_\star$.  These values have relative uncertainties of
$\approx$5\%, based on the uncertainties in the measured velocity
dispersions.  The two inner clumps are centered at orbital distances
of 2.07\,$R_\star$ and 2.88\,$R_\star$.  There is therefore physical
overlap in their spatial distributions.  These two clumps could in
fact be a single clump with an optically thick H$\alpha$ line core.
Regardless, the implication is that the full length in the radial
dimension of these two inner emitting clumps is approximately equal to
the star's diameter.  At mid-transit, these inner clumps have a
similar velocity dispersion, although with greater uncertainty due to
the differences between the $\phi$=0 and $\phi$=1 transits (see
Figure~\ref{fig:halphamodel}).  This suggests a 1$\sigma$ emission
contour in the azimuthal dimension with a length of
$\ell$$\approx$0.5\,$R_\star$.

The vertical height of the emitting region is less constrained because
the system is consistent with being viewed edge-on.  However, one
possible constraint follows by assuming that the TESS transit depth
scales with the projected H$\alpha$ emission area $\ell z$.  More
specifically, one can evaluate an ``effective area'' blocked by a
two-dimensional gaussian blob passing over a star by integrating the
local gaussian weight over the stellar surface.  For instance, if
$\ell$$\approx$$z$$\approx$0.24\,$R_\star$, then
$\iint \exp \left(
  -\left( x^2/2\sigma_x^2 + y^2/2\sigma_y^2 \right)
\right) {\rm d}x \, {\rm d}y$
suggests 11.5\% of the star being ``blocked'', a geometric factor
which would need to be in turn multiplied by an unknown opacity factor
to produce the observed transit depth ($\delta$$\approx$5\%).  There
is a degeneracy between $z$ and this opacity factor; larger vertical
heights are allowed for lower optical depths in absorption, and
vice-versa.  This constraint also implicitly assumes that the
optically thick material is well-mixed with the hydrogen, which may
not be accurate.

\section{Properties of the Plasma and Magnetospheric Environment}
\label{subsec:gas}

The hydrogen number density, plasma temperature, and magnetic field
strength inside the clumps can be estimated from the available data.  
The following order-of-magnitude calculations assume a simple
uniform-density plasma in a spherical geometry: more careful
considerations of radiative transfer are a worthy topic for future
work.

Circumstellar H$\alpha$ emission could come either from resonant
scattering of stellar H$\alpha$ photons, or from radiative
recombination.  We neglect scattering because Figure~\ref{fig:spec}
shows the circumstellar H$\alpha$ emission varying by a factor of
$\approx$5 while the chromospheric line core is stable.  The volume
emissivity under case B recombination can be written \begin{equation}
  j_{\rm H\alpha} = n_{\rm e} n_{\rm p} \alpha^{\rm eff}_{\rm H\alpha}
h \nu_{\rm H\alpha}, \end{equation} where $n_{\rm e}$ and $n_{\rm p}$
are the electron and proton densities, and $\alpha^{\rm eff}_{\rm
H\alpha}$ is the effective recombination coefficient, defined to
include all recombination routes that produce an H$\alpha$ photon.
For hydrogen with temperatures between 1,000-10,000\,K, $\alpha^{\rm
eff}_{\rm H\alpha}$ is typically $10^{-12}$ to
$10^{-13}$\,cm$^3$\,s$^{-1}$ \citep{Hummer1987,Draine2011}.
Neglecting atoms other than hydrogen, we can assume an ionization
fraction $x$, such that $n_{\rm e} = n_{\rm p} = x n_{\rm H}$, for
$n_{\rm H}$ the hydrogen number density.  Let $L_{\rm H\alpha} =
j_{\rm H\alpha} V$, for $V$ the volume of the emitting hydrogen.  The
luminosity of circumstellar hydrogen emission, $L_{\rm H\alpha}$, is
an observable: our SED fitting routine yields
$L_\star$$\approx$0.012\,$L_\odot$, which implies that the stellar
H$\alpha$ line radiates at $\approx$1.0$\times$$10^{28}$erg\,s$^{-1}$.
The luminosity of the clumps $L_{\rm H\alpha}$ are of order one tenth
that of the star.  If we approximate the emitting volume as a
homogeneous sphere of radius $r$, we can write
\begin{align}
  n_{\rm H} &= 
  1 \cdot 10^{11}\,{\rm cm}^{-3}
  \left(
    \frac{0.5}{x}
  \right)
  \left( 
    \frac{ L_{\rm H\alpha} }{ 10^{27}\,{\rm erg\,s^{-1}} }
    \frac{ 10^{-13}\,{\rm cm^{3}\,s^{-1}} }{ \alpha^{\rm eff}_{\rm H\alpha} }
  \right)^{1/2}
  \left(
    \frac{ 0.1\,R_\odot }{ r }
  \right)^{3/2}.
  \label{eq:numdensity}
\end{align}
For a uniform density clump, this suggests a total gas mass of $M_{\rm
gas}\approx 2\times10^{17}$\,g, which is similar to masses derived for
plasma clumps around other cool stars \citep{VillarrealDAngelo2019}.
We emphasize that Equation~\ref{eq:numdensity} in intended to provide
only an order of magnitude estimate for the number density implied by
the observed H$\alpha$ emission.  In detail, the effective
recombination rate and the ionization fraction each vary with density
and temperature; a more thorough estimate would iteratively solve the
equations of detailed balance and radiative transfer (e.g.~Figure~8 of
\citealt{CollierCameron1989}), and potentially also consider
departures from local thermodynamic equilibrium.

Finally, a constraint on the magnetic field strength at the site of
the clump follows from the requirement that the magnetic pressure
exceed the thermal pressure, $B_{\rm c}^2 / 8\pi > n_{\rm H} k T$.
Although we do not know the exact plasma temperature, if it were
significantly beyond 1,000-10,000\,K, we would either fully ionize the
hydrogen, or not ionize enough of it.  The field strength at the clump
must therefore exceed
\begin{equation}
  B_{\rm c} \gtrsim 1\,{\rm G}
  \left(
  \frac{n_{\rm H}}{1\times10^{11}\,{\rm cm}^{-3}}
  \frac{T}{3000\,{\rm K}}
  \right)^{1/2}.
\end{equation}
Given that the average surface magnetic field strengths of low-mass
stars have been measured to span hundreds to thousands of Gauss
\citep{Donati2009,Kochukhov2021,Reiners2022}, this bound is easily met
at orbital distances of 2-4\,$R_\star$.

\section{Upper and Lower Bounds on Dust}
\label{subsec:dust}

The material's composition -- either pure plasma, or a dusty plasma --
is not known.  The idea of dust being present seems plausible given
observations of chromatic transits in analogous objects
\citep{Tanimoto2020,Gunther2022,Koen2023}.  However, this scenario is
highly constrained.  An upper limit on the amount of hot dust follows
from the lack of an infrared excess.  A lower limit follows if one
assumes that most of the broadband optical depth comes from dust
absorption and scattering, rather than any radiative processes
associated with the plasma.

Regarding the upper limit, Figure~\ref{fig:sed} shows the SED.  While
AllWISE \citep{Cutri2014} yielded a confident W3 detection
(9.8$\sigma$) consistent with the photospheric extrapolation from
bluer bandpasses, the W4 extraction yielded only a marginal indication
(1.7$\sigma$) of detectable flux.  Similar to other CPVs
\citep{Stauffer2017,Bouma2024}, the photometric uncertainties from
WISE W1 and W2 allow at most a $\lesssim$2\% excess at 3-5\,$\mu$m
relative to the stellar photosphere, and a $\lesssim$5\% excess at
10\,$\mu$m (W3).  To estimate the implied mass bound, we assume a dust
temperature $T_{\rm d}$=1500\,K, typical for dust near the star (see
\citealt{Zhan2019} for discussion regarding dust sublimation).  We
then treat emission from the dust and star as Planck functions, and
require $L_{\rm d} < f L_\star$, where the factor $f$ is set by the
photometric precision of WISE and $L_{\rm d}$ is the bolometric dust
luminosity.  Given the reported uncertainties, we numerically find
$f<6\cdot$10$^{-3}$.  From the Stefan-Boltzmann law we can then write
$A_{\rm d} < f (T_\star / T_{\rm d})^4 Q_{\rm em}^{-1} (4\pi
R_\star)^2$, for $A_{\rm d}$ the total emitting surface area of the
dust, and $Q_{\rm em}$ an emission efficiency parameter.  Converting
this constraint to a dust mass requires an assumption regarding the
grain properties.  We assume a grain density $\rho_{\rm
d}$=3\,g\,cm$^{-3}$ typical for silicate grains, a fixed grain size
$a$=1\,$\mu$m, and no self-absorption.  This enables the assumption
that $A_{\rm d} = N \pi a^2$, for $N$ the total number of dust grains.
This in turn yields an upper limit on the dust mass of
\begin{equation}
  M_{\rm dust} \lesssim 4 \cdot 10^{17}\, {\rm g}\ 
  \left( \frac{f}{6\cdot10^{-3}} \right)
  \left( \frac{T_\star}{3000\,{\rm K}} \frac{1500\,{\rm K}}{T_{\rm d}} \right)^4
  \left( \frac{Q_{\rm em}}{1} \right)^{-1}
  \left( \frac{R_\star}{0.4\,R_\odot} \right)^2
  \left( \frac{a}{1\,\mathrm{\mu m}} \right)
  \left( \frac{\rho_{\rm d}}{3\,\mathrm{g\,cm^{-3}}} \right).
\end{equation}

The analogous lower limit follows from requiring the optical depth
from absorption and scattering $\tau$ to be at least unity.  The
optical depth can be written $\tau = n \sigma \ell$, where $\sigma$ is
the cross-section, $n$ is the number density, $\ell$ is the path
length.  For spherical dust grains in the optical, $\sigma = Q_{\rm
ext} \pi a^2$, where $Q_{\rm ext}$ is the extinction efficiency
parameter, tabulated e.g.~by \citet{Croll2014} in their Figure 13.
\citet{Sanderson2023} calculated the relevant cloud mass for this
problem assuming a spherical dust clump of size $r$, and they found
\begin{equation}
  M_{\rm dust} \gtrsim 2 \cdot 10^{15}\, {\rm g}\ 
  \left( \frac{\tau}{1} \right)
  \left( \frac{Q_{\rm ext}}{3} \right)^{-1}
  \left( \frac{r}{0.1\,R_\star} \frac{R_\star}{0.4\,R_\odot} \right)^2
  \left( \frac{a}{1\,\mathrm{\mu m}} \right)
  \left( \frac{\rho_{\rm d}}{3\,\mathrm{g\,cm^{-3}}} \right).
\end{equation}

Three relevant objects for comparison include solar prominences,
planetesimals, and comets.  Prominences of the Sun have gas masses of
$10^{14}$\,g-$10^{16}$\,g \citep{VialEngvold2015}.  A planetesimal of
mass $\approx$10$^{15}$\,g with a bulk density of 1\,g\,cm$^{-3}$
would have a diameter of order one kilometer.  Halley's comet has a
mass of order $10^{17}$\,g \citep{Rickman1989}, of which
$\sim$$10^{14}$\,g is shed per orbit, most of which inspirals toward
the Sun due to Poynting-Robertson drag.

To summarize, if dust is responsible for the broadband variability of
CPVs, it would need to be concentrated in clumps with masses in the
range of $10^{15}$-$10^{17}$\,g.  Given $M_{\rm
gas}$$\approx$2$\times$10$^{17}$\,g from Appendix~\ref{subsec:gas},
the allowed dust masses imply $M_{\rm gas}/M_{\rm dust}$ ranges of
1-100.  More careful measurements of this ratio---in particular by
inferring the dust mass through high precision infrared
spectrophotometry---could provide a path for distinguishing the
scenario of a trapped stellar outflow from an accumulation of
externally-sourced material.  While there are several plausible
external sources, feeding through a low-mass disk in particular cannot
be ruled out based on typical disk depletion times \citep{Haisch2001}.
Observations of infrared excesses and accretion signatures in low-mass
stars tens of millions of years old suggest a broad lifetime
distribution for such disks
\citep{Silverberg2020,Lee2020,Gaidos2022,Pfalzner2024}.

\begin{table}
  \footnotesize
  \centering
  \begin{tabular}{lcr}
  \hline 
  \hline 
  Time [BJD$_{\rm TDB}$] & RV (\kms) & $\sigma_{\rm RV}$ (\kms) \\
  \hline 
  2460357.954919 & 2.73 & 5.86 \\
  2460357.965845 & -4.40 & 2.37 \\
  2460357.976770 & -0.19 & 2.64 \\
  2460357.987698 & 3.84 & 2.87 \\
  2460357.998619 & 7.53 & 7.53 \\
  2460358.009538 & -1.98 & 1.44 \\
  2460358.020462 & 1.02 & 1.21 \\
  2460358.031383 & 0.64 & 7.03 \\
  2460358.042306 & -2.91 & 2.71 \\
  2460358.053228 & 8.93 & 6.75 \\
  2460358.064154 & 5.95 & 8.84 \\
  2460358.075075 & -2.25 & 3.06 \\
  2460358.085996 & 1.84 & 1.34 \\
  2460358.096918 & 2.41 & 8.24 \\
  2460358.107839 & -7.04 & 3.94 \\
  2460358.118760 & -2.24 & 3.07 \\
  2460358.129683 & -2.83 & 7.55 \\
  2460358.140606 & -0.59 & 2.26 \\
  2460358.151527 & 1.84 & 2.91 \\
  2460358.162448 & 4.54 & 3.95 \\
  2460358.173368 & 6.21 & 12.14 \\
  \hline
  \end{tabular}
  \caption{TIC\,141146667 radial velocities relative to the systemic
  velocity based on the 7699\,\AA\ resonance line and TiO bandheads.}
  \label{tab:rv}
\end{table}

\bibliographystyle{aasjournal}
\bibliography{cpvbib.bib}


\end{document}